\documentclass[twocolumn,prl,superscriptaddress,showpacs]{revtex4}
\usepackage{graphicx}
\usepackage{dcolumn}
\usepackage{bm}
\usepackage{amsmath}

\begin{document}

\preprint{}
\title{Chirality sensitive effect on surface states in chiral $p$-wave superconductors}
\author{Takehito Yokoyama}
\affiliation{Department of Applied Physics, Nagoya University, Nagoya, 464-8603, Japan\\
and CREST, Japan Science and Technology Corporation (JST) Nagoya, 464-8603, Japan}
\author{Christian Iniotakis}
\affiliation{Institute for Theoretical Physics, ETH Zurich, 8093 Zurich, Switzerland}
\author{Yukio Tanaka}
\affiliation{Department of Applied Physics, Nagoya University, Nagoya, 464-8603, Japan\\
and CREST, Japan Science and Technology Corporation (JST) Nagoya, 464-8603, Japan}
\author{Manfred Sigrist}
\affiliation{Institute for Theoretical Physics, ETH Zurich, 8093 Zurich, Switzerland}
\date{\today}

\begin{abstract}
We study the local density of states at the surface of a chiral $p$-wave superconductor in the presence of 
a weak magnetic field. As a result, the formation of low-energy Andreev bound states is either suppressed
or enhanced by an applied magnetic field, depending on its orientation with respect to the chirality of the $p$-wave superconductor.
Similarly, an Abrikosov vortex, which is situated not too far from the surface, leads to a zero-energy peak
of the density of states, if its chirality is the same as that of the superconductor, and to a gap structure for the opposite case. 
We explain the underlying principle of this effect and propose a 
chirality sensitive test on unconventional superconductors. 
\end{abstract}

\maketitle



%

%



Much attention has been paid to unconventional superconductors, because they can exhibit a sign or general phase change of their gap function as a function of momentum. This property induces many intriguing phenomena, which can be observed directly by so-called phase sensitive experiments providing powerful tools  to test the symmetry of the gap function \cite{Sigrist}. 
One important consequence of the sign change of the gap function is the possible existence of Andreev bound states at 
the surface of the superconductor
\cite{Hu,Tanaka,Buchholtz}.
The formation of Andreev bound states increases the local zero-energy quasiparticle density of states (DOS) 
at the surface, leading to a pronounced zero-bias conductance peak
in the tunneling conductance observable both in singlet $d$-wave superconductors like
the cuprates and in triplet $p$-wave superconductors such as Sr$_2$RuO$_4$
\cite{Lesueur,Covington,Review,Yamashiro,Honerkamp,Matsumoto,Laube,Mao,Yokoyama}. 
For the case of $d$-wave superconductors it is well-known, that an applied magnetic field 
or an applied electric current result in a split of this zero-bias conductance
peak, since the zero-energy spectral weight of the bound states is effectively Doppler shifted 
towards higher energies \cite{Fogel,Covington,Aprili,Dagan}.
The same effect also appears 
for an Abrikosov vortex, which is pinned not too far from the boundary. Here,
the zero-energy DOS is suppressed in a shadow-like region 'behind' the vortex \cite{Graser,Iniotakis}.

Regarding the chiral $p$-wave superconducting phase as it is likely realized in Sr$_2$RuO$_4$,
a further aspect appears. The chiral $p$-wave state characterized by the  vector $\bm{d}(\bm{k}) = (0,0,k_x  \pm ik_y)$ breaks time reversal symmetry \cite{Maeno,Ishida,Luke,Nelson,Mackenzie}. 
In this Letter we will show that the influence of an external magnetic field on the surface density of states is selective for the chirality. The quasiparticle density of states at the surface increases or decreases depending on the relative orientation of the applied magnetic field and the chirality. Similarly, we find that the influence of a vortex on the surface states depends on the orientation of vorticity with respect to chiralty. These characteristic
effects could open an alternative way to chirality sensitive probes, in contrast to phase or spin sensitive setups, which have been 
intensively used in the scientific community already.


For our calculations, we use quasiclassical Eilenberger theory of superconductivity 
\cite{Eilenberger,Larkin}
in the so-called Riccati-parametrization \cite{SchopohlMaki}, which allows to achieve numerically stable 
solutions for the quasiclassical propagators (and thus also for the DOS) in spatially nonhomogeneous systems.
Concretely, in our case we consider a superconducting half space $x>0$ exhibiting
a gap function of chiral $p$-wave symmetry. The surface at
$x=0$ is included in our calculations in a straightforward way by specific boundary conditions 
\cite{Zaitsev, Shelankov, Eschrig}. 
We assume a cylindrical 
Fermi surface for the superconductor with the symmetry axis pointing
along the $z$-direction, so that the Fermi velocity can be 
parametrized by the polar angle $\theta$ via 
${\bm v}_F = v_F ({\bm {\hat{x}}} \cos \theta +{\bm {\hat{y}}} \sin \theta)$.
For a given point ${\bm r}_0$ in space and angle $\theta$ parametrizing the
Fermi surface, a quasiclassical trajectory is then defined according to 
${\bm r}(x') = {\bm r}_0 + x' \; \hat{\bm {v}}_F $. Along such a trajectory
the Eilenberger equations can be transformed into $2\times 2$ matrix differential equations in spin
space, which are of the  Riccati type and can be solved much easier \cite{Eschrig}.
In our case, we deal with a one-component gap-function, so that the corresponding
Riccati equations take the simpler form \cite{SchopohlMaki,Matsumoto}
\begin{eqnarray}
\hbar v_F \partial_{x'}  a(x') + \left[ 2 \tilde{\epsilon}_n + \Delta^\dagger a(x') \right] a(x')
- \Delta & = & 0 \nonumber \\
\hbar v_F \partial_{x'} b(x') - \left[ 2 \tilde{\epsilon}_n + \Delta b(x') \right] b(x')
+ \Delta^\dagger & = & 0 \label{Riccati}
\end{eqnarray}
for the two scalar coherence functions $a$ and $b$. Here, 
$i \tilde{\epsilon}_n = i \epsilon_n + {\bf v}_F \cdot \frac{e}{c} {\bm A}$ 
denotes Matsubara frequencies which are shifted due to the presence
of a magnetic vector potential $\bm A$.
The pairing potential $\Delta$ can be factorized 
in the following form
\begin{equation}
\Delta ({\bm r}, \theta) = \Delta_0 \exp( i \theta ) \Psi ({\bm r}). \label{delta}
\end{equation}
Here, $\Psi$ denotes a factor which covers the spatial dependence
of the pairing potential in general. Since we are only interested in the
main qualitative aspects of the local DOS, namely, if the zero-energy
spectral weight at the surface is suppressed or increased, we may take the
modulus of $\Psi$ to be constant \cite{Hu,Graser,Iniotakis}.
For the calculation of physical properties, the Riccati equations (\ref{Riccati}) have 
to be integrated numerically using proper starting values in the bulk.
The local DOS, which is already normalized to the DOS 
in the normal state, is then achieved by an integration over the Fermi
surface. In terms of the coherence functions $a$ and $b$, we have  
\begin{equation}
N({\bm r}_0,E) = \int_0^{2\pi} \frac{d \theta}{2 \pi}  {\rm Re} \;
\left[ \frac{1-a  b}{1+ a b} \right]_{i \epsilon_n \rightarrow E + i \delta},
\label{angularaverage}
\end{equation}
where $E$ denotes the quasiparticle energy with respect to the Fermi level and 
$\delta$ is an effective scattering parameter that corresponds to an inverse 
mean free path. For all numerical calculations, we fix this value as $\delta=0.1\Delta_0$.


In order to study the basic effect of chirality we consider a magnetic field applied along the $z$-axis at the surface, represented by a nearly homogeneous vector potential $\bm{A}$. 
We choose the real gauge, i.e. the spatially dependent part $\Psi$ of the pairing
potential is taken to
be real.
Since we additionally assumed a spatially constant modulus of the pairing potential, it is possible
to get analytical solutions for the coherence functions $a$ and $b$ in this case, which also
allows to examine the corresponding behaviour of the local DOS analytically.
Directly at the surface, we get
\begin{eqnarray}
N(E) = 2{\mathop{\rm Re}\nolimits} \left\langle {\frac{1}{{1 + a_{in} b_{out} }}} \right\rangle _{i \epsilon_n \rightarrow E + i \delta} - 1
\end{eqnarray}
with $a_{in}  = s\Delta _0 e^{i (\pi-\theta) }, b_{out}  = s\Delta _0 e^{-i \theta }$ and the abbreviation $s =1/(\tilde \varepsilon _n  + \sqrt {\tilde \varepsilon _n^2  + \Delta_0^2 })$. Furthermore, $\left\langle...\right\rangle$ denotes angular averaging, which we may restrict
to outgoing angles $-\pi/2 \leq \theta \leq \pi/2$ only. 
This directly yields
\begin{eqnarray}
N(E) = 2{\mathop{\rm Re}\nolimits} \left\langle {\frac{1}{{1 -(1 - 2\tilde \varepsilon _n s)e^{ - 2i\theta } }}} \right\rangle _{i \epsilon_n \rightarrow E + i \delta} - 1.
\end{eqnarray}
 Expanding the zero-energy DOS in orders of the vector potential $\bm A$, we obtain in the clean limit of $\delta \to 0^+$
\begin{eqnarray}
 N(E=0) = 1 + \frac{{ev_F }}{{c\Delta_0 }}A_y  +... \label{EqN}
\end{eqnarray}
Physically, this result displays the influence of a Doppler shift due to 
a superfluid velocity on the local quasiparticle spectrum. In terms of chiral
surface states \cite{ChiralSurface,Furusaki}, the Doppler shift leads to a change of the slope of the 
quasiparticle dispersion [$\epsilon(k_y)=\Delta_0 k_y/k_F $], which has a direct impact on the
corresponding DOS.
As is seen from the result on the right-hand side of  Eq. (\ref{EqN}), the term of the vector potential, 
which belongs to the direction perpendicular 
to the chirality of the $p$-wave superconductor, survives in linear order. This is in contrast
to other superconductors like  $s$-, $d$- or $p$-wave superconductors without chirality, 
where we obtain  similarly
\begin{equation}
N(E=0) = C + \left\langle F (\theta) \sin\theta \right\rangle A_y  + ...
\end{equation}
with a constant $C$ and a  function $F$ which satisfies $F (\theta) = F (-\theta)$. 
Thus, after angular averaging, the coefficient of the linear term vanishes in these cases, 
reflecting the presence of inversion symmetry with respect to the x-y plane. 
Since an applied magnetic field is related to the vector potential as
$B_z  =  \partial_x A_y$, 
the zero-energy DOS in the chiral $p$-wave superconductor depends on the sign of the magnetic field. 
Applying a weak magnetic field along the chirality direction suppresses the 
zero-energy bound states, while applying it in the opposite direction leads 
to a zero-energy peak of the surface DOS. 
It is important to realize that the derived  Eq. (\ref{EqN}) qualitatively implies this chirality 
sensitive effect also for the case of a more general vector potential.
Especially, this chirality effect is remarkable in the presence of a vortex near the surface.


\begin{figure}[t]
\includegraphics[width= 0.99 \columnwidth]{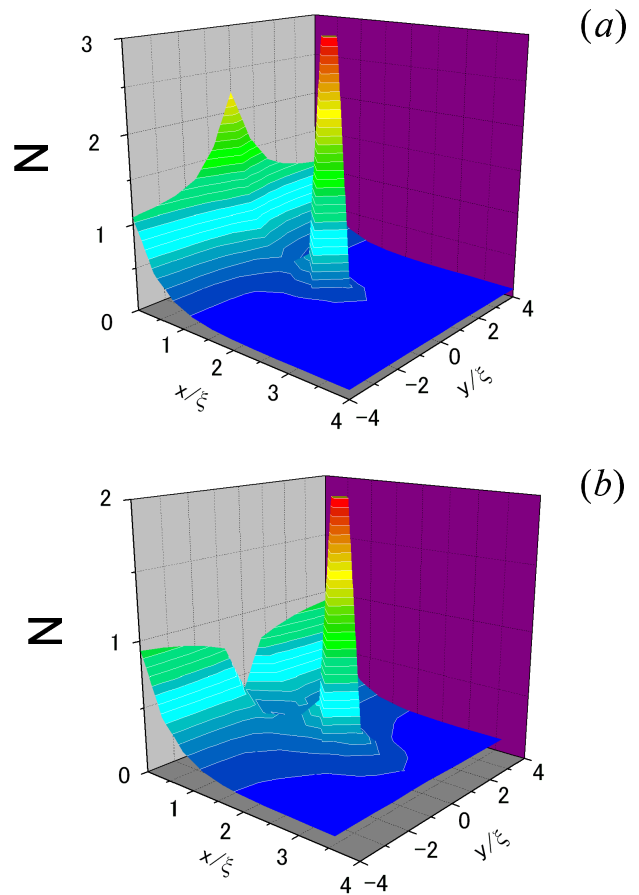}
\caption{(color online) $N(E=0)$ in a chiral $p$-wave superconductor in the presence of a vortex, which is situated at a distance of
$x_V=2\xi$ from the surface. The vortex and the $p$-wave state have (a) the same chirality and (b) the opposite chirality. Clearly, there is a 
remarkable difference in the quasiparticle spectral weight at the surface ($x=0$) behind the vortex, showing a strong increase in (a) and a suppression in (b).} \label{f1}
\end{figure}

\begin{figure}[t]
\includegraphics[width=0.99 \columnwidth]{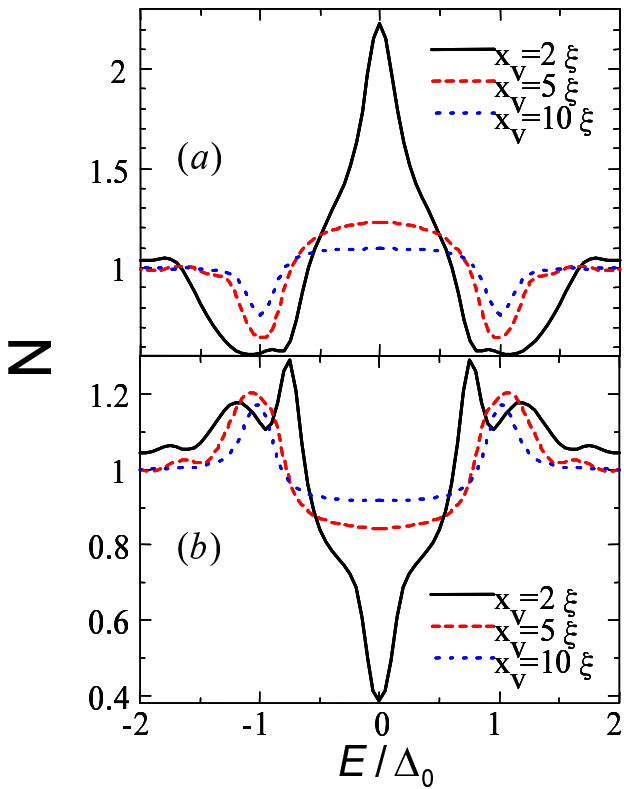}
\caption{(color online) Local DOS at the
point $x=y=0$ for different vortex to boundary distances $x_V$ as
a function of energy. Again, $p$-wave state and Abrikosov vortex have the same chirality in
(a) and the opposite in (b), resulting in peak and dip structures, respectively, around zero energy. } \label{f3}
\end{figure}

As a next example of the chirality effect, we study the case of a single  Abrikosov vortex line parallel
to the $z$-axis. For such a vortex in the bulk of a chiral $p$-wave superconductor,  the interplay 
between vorticity and chirality has been intensively studied already \cite{Hayashi}. In the present work, however, we 
focus on important surface effects, which appear at the boundary due to the presence of the vortex.
We assume the vortex to be at a distance $x_V$ from the boundary. With $y$ denoting the coordinate along
the boundary, the vortex shall sit at $y_V=0$. 
For convenience, we set the vector potential $\bm{A}=0$ and incorporate the
vortex  by a properly chosen  phase factor  
$\Psi ({\bm r}) =  e^{i \Phi({\bm r})}$ instead,
which (in standard complex notation $z=x+iy$) is given by \cite{Graser, Iniotakis}
\begin{equation}
e^{i \Phi({\bm r})}=\frac{z-z_V}{|z-z_V|}\cdot \left( \frac{z-\bar{z}_V}{|z-\bar{z}_V|} \right)^*. \label{ePhi}
\end{equation}
Here, the first factor is the phase of a single vortex at position $z_V$, while the second
factor corresponds to a virtual antivortex placed at the mirrored position $\bar{z}_V=-x_V$,
which  ensures the correct implementation of boundary conditions.
In the following, we consider two different cases: The chiralities of the $p$-wave state and the vortex are the 
same (a), or  opposite (b). The latter case is implemented by a complex conjugation of the 
phase factor given in Eq. (\ref{ePhi}), replacing the vortex by an antivortex and vice versa. 
Our results for the corresponding local zero-energy DOS near the surface of the $p$-wave superconductor 
are shown in Fig. \ref{f1}. The Abrikosov vortex position is fixed  at a
distance of $x_V=2\xi$ from the boundary with $\xi=\hbar v_F/\Delta_0$ denoting the 
coherence length.
Apart from zero-energy bound states in the vortex core, there are also bound states at
the surface of the $p$-wave superconductor. Far away from the vortex, these surface
bound states have the spectral weight of $N(E=0)=1$ [cf. Eq. (\ref{EqN})]. However, as can be seen quite
clearly in Fig. \ref{f1}, the local DOS drastically changes in a shadow-like region behind 
the vortex. If vortex and $p$-wave state have the same chirality, the bound states are
strongly enhanced (a), for opposite chirality they are suppressed (b). The latter 
suppression resembles a similar effect appearing in $d$-wave superconductors \cite{Graser, Iniotakis}.

In Fig. \ref{f3}, we show the local  DOS at the 
point $x=y=0$ for different vortex to boundary distances $x_V$ as
a function of energy. Around zero energy, we find a sharp peak or dip structure,
respectively,  again depending on the chirality. These structures get
less pronounced, when the vortex distance is increased, nevertheless
they persist. Moreover, the quasiparticle spectrum starts to exhibit some kind of 
mirror symmetry around the value $N(E)=1$ for the two different chiralities.
Note that this is in qualitative  agreement
with Eq. (\ref{EqN}) since after a transformation to the real gauge 
we eventually have $A_y>0$ due to the vortex, leading  to
the strong increase of Andreev bound states at the surface for the same
chirality. The result for opposite chirality  is obtained due to $A_y<0$ for the antivortex, accordingly.
It is worth noting that the modification of the surface quasiparticle states due to the presence of vortices has an effect on the force acting on vortices near the surface. An increase of the DOS leads to a repulsion of the vortex from the boundary towards the bulk, whereas 
a decrease results in an attraction towards the boundary. Thus, in both cases the Bean-Livingston barrier would be modified, which influences the escape and entrance of vortices to the superconductor \cite{Bean, IniotakisBean}.

Our results allow us to propose a chirality sensitive test on superconductors, 
based on well-established experimental techniques, which are capable of indicating  the
weight of Andreev bound  states, for example tunneling conductance experiments or  low-temperature scanning tunneling
spectroscopy \cite{Fischer}.
For a chiral superconductor, it is expected to observe a suppression of the zero-energy DOS at the 
surface, when a weak magnetic field is applied parallel to the chirality. Inverting the field, however, 
leads to an enhancement of the DOS. In this way chirality could be detected. This unusual reversal effect does not appear in non-chiral superconductors.
Moreover, the experiment allows us to detect the chirality and possibly even the domains of
different chirality, if domain walls reach the surface. 


In summary, we have studied the DOS at the surface of a chiral $p$-wave superconductor
in the presence of a magnetic vector potential due to, for example, an applied 
magnetic field or an  Abrikosov vortex. We clarified that the weight of low-energy surface bound states gets either suppressed or increased, 
depending on the orientation of both chirality and magnetic field. Due to this chirality-sensitive effect on the Andreev
bound states, a setup to test the chirality of an unconventional superconductor could be accessible experimentally.
Furthermore,  for vortices this effect also has  a  chirality selective
influence on the Bean-Livingston barrier which could give rise to a different escape rate of vortices from the two kinds of 
chiral domains.


%

This study was supported by the Japan Society for the Promotion of Science (T.Y.), Grant-in-Aid for Scientific Research Grant No. 17071007 on Priority Area "Novel Quantum Phenomena Specific to Anisotropic Superconductivity"  (T.Y. and Y.T.) as well as the Swiss Nationalfonds and the NCCR MaNEP
(C.I. and M.S.).
%



\begin{thebibliography}{99}

\bibitem{Sigrist} M. Sigrist and K. Ueda, Rev. Mod. Phys. \textbf{63}, 239 (1991). 

\bibitem{Hu} C.~R.~Hu, {Phys.~Rev.~Lett.} {\bf 72}, 1526 (1994). 
\bibitem{Tanaka} Y.~Tanaka and S.~Kashiwaya, {Phys.~Rev.~Lett.} {\bf 74}, 3451 (1995). 
\bibitem{Buchholtz} L.~J.~Buchholtz, M.~Palumbo, D.~Rainer, and J.~A.~Sauls, {J.~Low~Temp.~Phys.} {\bf 101}, 1099 (1995).

\bibitem{Lesueur} J.~Lesueur, L.~H.~Greene, W.~L.~Feldmann, and A.~Inam, {Physica C} {\bf 191}, 325 (1992).
\bibitem{Covington} M. Covington, M. Aprili, E. Paraoanu, L.H. Greene, F. Xu, J. Zhu, and C.A. Mirkin, 
Phys. Rev. Lett. \textbf {79}, 277 (1997).
\bibitem{Review}S. Kashiwaya and Y. Tanaka, Rep. Prog. Phys. \textbf{63}, 1641 (2000). 

\bibitem{Yamashiro} M. Yamashiro, Y. Tanaka, and S. Kashiwaya, Phys. Rev. B  \textbf{56}, 7847 (1997).
\bibitem{Honerkamp} C. Honerkamp and M. Sigrist, J. Low Temp. Phys. {\bf 111}, 895 (1998).
\bibitem{Matsumoto} M. Matsumoto and M. Sigrist, J. Phys. Soc. Jpn. {\bf 68}, 994 (1999).
\bibitem{Laube} F. Laube, G. Goll, H. v. L\"ohneysen, M. Fogelstr\"om, and F. Lichtenberg, 
Phys. Rev. Lett. {\bf 84}, 1595 (2000).
\bibitem{Mao} Z.Q. Mao, K.D. Nelson, R. Jin, Y. Liu, and Y. Maeno, Phys. Rev. Lett. {\bf 87}, 037003 (2001).

\bibitem{Yokoyama} T. Yokoyama, Y. Tanaka, and J. Inoue, Phys. Rev. B  \textbf{72}, 220504(R) (2005).

\bibitem{Fogel} M.~Fogelstr\"om, D. Rainer, and J. A. Sauls, Phys. Rev. Lett. \textbf {79}, 281 (1997). 
\bibitem{Aprili} M.~Aprili, E.~Badica, and L.~H.~Greene, {Phys.~Rev.~Lett.} {\bf 83}, 4630 (1999).
\bibitem{Dagan} Y.~Dagan and G.~Deutscher, {Phys.~Rev.~Lett.} {\bf 87}, 177004 (2001).

\bibitem{Graser} S. Graser, C. Iniotakis, T. Dahm, and N. Schopohl, Phys. Rev. Lett. {\bf 93}, 247001 (2004)
\bibitem{Iniotakis} C. Iniotakis, S. Graser, T. Dahm, and N. Schopohl, Phys. Rev. B {\bf 71}, 214508 (2005)

\bibitem{Maeno} Y. Maeno, H. Hashimoto, K. Yoshida, S. Nishizaki, T. Fujita, J. G. Bednorz, and F. Lichtenberg, 
Nature (London) \textbf{372}, 532 (1994).
\bibitem{Ishida} K. Ishida, H. Mukuda, Y. Kitaoka, K. Asayama, Z. Q. Mao, Y. Mori, and Y. Maeno, 
Nature (London) \textbf{396}, 658 (1998).
\bibitem{Luke} G. M. Luke, Y. Fudamoto, K. M. Kojima, M. I. Larkin, J. Merrin, B. Nachumi, Y. J. Uemura, Y. Maeno, 
Z. Q. Mao, Y. Mori, H. Nakamura, and M. Sigrist, Nature (London) \textbf{394}, 558 (1998).
\bibitem{Mackenzie} A. P. Mackenzie and Y. Maeno, Rev. Mod. Phys. \textbf{75}, 657 (2003).
\bibitem{Nelson} K. D. Nelson, Z. Q. Mao, Y. Maeno, and Y. Liu, Science \textbf{306}, 1151 (2004); Y. Asano, Y. Tanaka, M. Sigrist, and S. Kashiwaya, Phys. Rev. B \textbf {67}, 184505 (2003); Phys. Rev. B \textbf {71}, 214501 (2005). 

\bibitem{Eilenberger} G.~Eilenberger, {Z.~Phys.} {\bf 214}, 195 (1968). 
\bibitem{Larkin} A.~I.~Larkin and Yu.~N.~Ovchinnikov, {Zh.~Eksp.~Teor.~Fiz.} {\bf 55}, 2262 (1968) 
[{Sov.~Phys.~JETP} {\bf 28}, 1200(1969)].  
\bibitem{SchopohlMaki} N.~Schopohl and K.~Maki, {Phys.~Rev.~B} {\bf 52}, 490 (1995);
N.~Schopohl, cond-mat/9804064.

\bibitem{Zaitsev} A.~V.~Zaitsev, {Zh.~Eksp.~Teor.~Fiz.} {\bf 86}, 1742 (1984) 
[{Sov.~Phys.~JETP} {\bf 59}, 1015 (1984)].
\bibitem{Shelankov} A.~Shelankov and M.~Ozana, {Phys.~Rev.~B} {\bf 61}, 7077 (2000).
\bibitem{Eschrig} M. Eschrig, Phys. Rev. B {\bf 61}, 9061 (2000).

\bibitem{ChiralSurface} M.~Sigrist, A.~Furusaki, C.~Honerkamp, M.~Matsumoto, K.-K.~Ng, and
Y.~Okuno, J. Phys. Soc. Jpn.  Suppl. B. {\bf 69}, 127 (2000).
\bibitem{Furusaki} A.~Furusaki, M.~Matsumoto, and M.~Sigrist, Phys. Rev. B {\bf 64}, 054514 (2001).

\bibitem{Hayashi} N. Hayashi and Y. Kato, Phys. Rev. B \textbf{66} 132511 (2002); N. Hayashi and Y. Kato, J. Low Temp. Phys. \textbf{131} 893 (2003).

\bibitem{Bean} C.~P.~Bean and J.~D.~Livingston, Phys. Rev. Lett. {\bf 12}, 14 (1964).
\bibitem{IniotakisBean} C.~Iniotakis, T.~Dahm, and N.~Schopohl, cond-mat/0705.1819.

\bibitem{Fischer} \O. Fischer, M. Kugler, I. Maggio-Aprile, C. Berthod, and C. Renner, Rev. Mod. Phys. \textbf{79}, 353 (2007).





\end{thebibliography}
\end{document}